# Simulation of Many-Body Fermi Systems on a Universal Quantum Computer

Daniel S. Abrams

Department of Physics, Massachusetts Institute of Technology, Cambridge, MA 02139 (abrams@mit.edu)

Seth Lloyd

Department of Mechanical Engineering, Massachusetts Institute of Technology,

Cambridge, MA 02139 (slloyd@mit.edu)

*Abstract*

We provide fast algorithms for simulating many body Fermi systems on a universal quantum computer. Both first and second quantized descriptions are considered, and the relative computational complexities are determined in each case. In order to accommodate fermions using a first quantized Hamiltonian, an efficient quantum algorithm for anti-symmetrization is given. Finally, a simulation of the Hubbard model is discussed in detail.

Since the discovery by Shor of a quantum algorithm for factoring in polynomial time [1], there has been tremendous activity in the field of quantum computation [2-9,11-18,20-22,26-27,32-33]. Recent results include the first experimental demonstrations of working quantum logic gates[2-4], quantum error-correcting codes [5], and many novel proposals for the design of actual quantum computers [3,6-10]. For a review of progress in quantum computation, see references 11 or 12. Despite these advances, the technical hurdles that stand in the way of factoring a large number on a quantum computer remain daunting[13-16]. But the problem of simulation - that is, the problem of modeling the full time evolution of an arbitrary quantum system - is less technologically demanding. While thousands of qubits and billions of quantum logic operations are needed to solve classically difficult factoring problems[17], it would be possible to use a quantum computer with only a few tens of qubits and a few thousand operations to perform simulations that would be classically intractable[18]. A quantum computer of this scale appears to be a realistic possibility.

Because the size of the Hilbert space grows exponentially with the number of particles, a full quantum simulation demands exponential resources on a classical computer. A system of only 100 spin 1/2 particles, for example, requires $2^{100}$ complex numbers to merely describe a general spin state. It is clear that on a classical computer, a simulation of this system is in general intractable. The idea that a quantum computer might be more efficient than a classical computer at simulating real quantum systems was first proposed by Feynman [19], but he speculated that the problem of Fermi statistics might prevent the design of a universal quantum simulator. More recently, Lloyd has shown how a quantum computer is in fact an efficient quantum simulator [18]; other work on simulations can be found in [20-22]. In this letter, we deal explicitly with the problem of fermions, in part by describing a quantum algorithm for antisymmetrization which executes in polynomial time. We also describe algorithms for quantum simulation of the Hubbard model in both first and second quantized formalisms. We show how the computer can be prepared in a state analogous to the initial state of a many body Fermi system, how it can be programmed to simulate the system's time evolution, and how measurements can then be made on the computer to extract relevant information, including, for example, the system's charge density distribution, two-point correlation functions, fluctuations, and scattering amplitudes. Thus this paper provides for the first time a complete quantum algorithm for simulating a system of physical interest, and describes the only known algorithm, other than Shor's, that gains an exponential speed increase by exploiting quantum computation (excepting certain artificial problems constructed explicitly for this purpose [23-25]).

The quantum computer used to perform the simulation could rely on a variety of possible physical systems to store and process quantum information: for example, photons which interact via small cavity Q.E.D. effects, electron spins, two level atoms interacting through NMR, or trapped ions [2,3,6,7]. The actual implementation of the quantum computer is not relevant, as long as it supports universal quantum computation [8,26-33] (although different physical implementations may of course be better or worse suited for different problems). Depending on the precise way in which the states of the physical system are mapped into the states of the computer, different algorithms will be needed to accomplish each of the three distinct stages of simulation: preparation, time evolution, and measurement. The problem that we consider here consists of n particles, each of which can be in any of m single particle states, labeled 1..m. These states might be sites in a lattice, or atomic orbitals, or plane waves, etc. The mapping of the model system onto the qubits of the computer depends on whether we choose a first or second quantized description. In many respects, the second quantized form appears naturally well-suited for quantum computation of Fermi systems: the occupation of each state must be either 0 or 1, which maps directly to the state of a quantum bit, or qubit [34]. In this case, the memory needed to map the state of the entire n particle system is m qubits (independent of n) [35]. To treat a first quantized Hamiltonian, we imagine a quantum word, or qu-word, as a string of qubits of length $\log_2 m$; one qu-word represents any integer in the range 1..m, and, consequently, the state of one particle. The state of the entire physical system being simulated can therefore be represented by n qu-words, or $n \log_2 m$ bits. If the number of particles is much smaller than the number of possible states, a first quantized representation may be vastly more efficient ($n \log_2 m$ qubits vs. m qubits). In either representation, if the simulated physical system is in a superposition of many direct product states (as it is in general), then the quantum computer will be in a corresponding superposition of states in order to represent the correct physical state of the simulated system.

The problem of fermions is handled more easily in the second quantized form. The calculation begins with all qubits in the |0> state. We can prepare the system in any state as long as it can be reached from the zero state using a relatively small number of quantum logic operations (that is, polynomial in m). Examples of such states include those in which the n particles are localized in individual lattice sites, momentum eigenstates, thermal states of non-interacting particles, and states in which particles obey k-particle correlations or entanglements (for small k). Thus Fermi statistics do not pose any additional complications for system preparation in the second quantized formalism. Because the statistics are incorporated into the raising and lowering operators, the additional complications occur during time-evolution. As a concrete example, we consider the Hubbard model, in which electrons move about a lattice of sites. Each site may be empty or occupied by a spin up electron, a spin down electron, or two electrons of opposite spin. Two qubits are therefore required to represent the four possible states of each site. The Hamiltonian for the system is

$$H = \sum_{i=1}^{m} V_0 n_{i\uparrow} n_{i\downarrow} + \sum_{<i,j>\sigma} t_0 c_{i\sigma}^* c_{j\sigma} \tag{1}$$

In the first term, which corresponds to potential energy, $V_0$ is the strength of the potential, and $n_{i\sigma}$ is the operator for the number of fermions of spin $\sigma$ at site i. In the second term, which corresponds to kinetic energy, the sum <i,j> indicates all neighboring pairs of sites, $t_0$ is the strength of the "hopping", and $c_{i\sigma}, c_{i\sigma}^*$ are operators for annihilation and creation, respectively, of a fermion at site i and spin $\sigma$. The computer simulates the Hubbard model by performing the unitary operation U = exp[(-i/h) H t] on suitably encoded states. This can be accomplished most easily by splitting the Hamiltonian into a sum of local terms $H_i$ and repeatedly applying the operators $U_i$ = exp[(-i/h) $H_i$ (t/n)], to evolve local parts of the system over small time slices t/n, in series. (See Ref. [18] for a detailed discussion of this technique). Thus it suffices to describe algorithms which perform the time-evolution corresponding to each local term in the Hamiltonian. To effect the time evolution corresponding to the potential energy terms $V = \sum_{i=1}^{m} V_0 n_{i\uparrow} n_{i\downarrow}$, the following simple algorithm will suffice: consider each site one at a time; for each site, if it is occupied by two electrons (of opposite spin), advance the phase of the entire state by [(-i/h) $V_0$ (t/n)]. This subroutine requires O(m) operations.

To calculate the effect of the hopping terms $\sum_{<i,j>\sigma} t_0 c_{i\sigma}^* c_{j\sigma}$ requires a slightly more complicated algorithm: first, we loop over all pairs of (physically) neighboring sites i and j, and consider hopping between each pair of sites separately. For each pair i,j, we count the number of occupied states which fall between i and j when the system is written in second quantized form. A flag is set to the parity of this number, which indicates whether or not a change of sign is introduced when hopping between the two sites, as required by the definitions of the raising and lowering operators. It is now easy to perform the time evolution $U_i$ corresponding to the Hermitian piece of the Hamiltonian $T_i = t_0(c_{j\sigma}^* c_{k\sigma} + c_{k\sigma}^* c_{j\sigma})$ by simply diagonalizing the Hamiltonian in the two qubit space i,j, and advancing the phase of the eigenstates.

Assuming that the number of neighbors is a constant, the loops execute O(m) times. It takes no more than O(m) operations to count the occupancy of the intervening states, and it follows that the entire algorithm for simulating the second quantized Hubbard model executes in $O(m^2)$ quantum logic operations.

Fermi statistics are more difficult to handle in the usual first quantized description, because it is necessary to initialize the quantum computer into an antisymmetrized superposition of states corresponding directly to the actual physical state of the system. As there are n! states in the superposition, one needs a fast quantum algorithm for generating this superposition state in order for the approach to be tractable. The algorithm we describe accepts as input a string of n qu-words (representing the state of the physical system being modeled) and generates an antisymmetrized superposition of n! states in $O(n^2(\ln m)^2)$ time. Note that without further restriction, antisymmetrization is an irreversible process and cannot be performed by a reversible quantum computer: there are n! input states which correspond to the same antisymmetrized state (modulo an overall phase) We therefore add the additional requirement that the input state must be ordered; i.e., that the number representing the state of the i'th particle is less than that of particle i+1, for all i<n-1. The correspondence between an ordered n-tuple of qu-words and an antisymmetrized superposition is one to one. In fact, this observation is in some sense the key to the algorithm.

System preparation in the first quantized formalism therefore begins by first initializing the computer into an unsymmetrized state and then antisymmetrizing that state. The system can be easily prepared in any (unsymmetrized) direct product state by merely placing each particle in the appropriate single particle state. These single particle states include, for example, those which are localized in position space, momentum space (obtained with a quantum FFT), and thermal states. The system can also be initialized into states with arbitrary k-particle correlations or entanglements by performing quantum logic operations in the appropriate k-particle space, requiring only $O(m^{2k})$ operations in the general case, and often far fewer.

Antisymmetrization is accomplished in four main steps, summarized below. A more detailed description will be published elsewhere [36]. *STEP I. Initialization of the input state.* We imagine that there is a string of qubits which are all initially set to zero, and define three registers A, B, and C, each consisting of n qu-words (n $\log_2$ m qubits). The qubits in register A are initialized to the ordered string of qu-words which represent the input state of the system $|\Psi\rangle$. The algorithm is unaffected if this state is a superposition of several ordered n-tuples. *Step II. Generating n! states.* We begin by creating the following superposition of states in register B:

$$\frac{1}{n!}\left(\sum_{1}^{n}|i\rangle\right)\otimes\left(\sum_{1}^{n-1}|i\rangle\right)\otimes\left(\sum_{1}^{n-2}|i\rangle\right)\otimes\ldots\otimes(|2\rangle+|1\rangle)\otimes|1\rangle \qquad (2)$$

This is accomplished with $O(n (\ln m)^2)$ steps: by performing appropriate rotations on each qubit, one at a time, the computer is placed in a superposition of n! unique states [37].

*Step III. Transform into permutations of natural numbers.* The goal of this third step is to transform the superposition of states in register B into the superposition $\frac{1}{n!}\sum_{\sigma \in S_n}|\sigma(1...n)\rangle$, where $S_n$ is the symmetric group of permutations on n objects. This is an equal superposition of the states representing all the permutations of the first n natural numbers. The basic idea is as follows: let B[i] indicate the i'th qu-word in register B; map B[i] into a qu-word B'[i] by setting B'[1] = B[1], and B'[i] equal to the B[i]th natural number which is not contained in the set {B'[1],...,B'[i-1]}, for i > 1. This transformation is effected with $O(n^2 \ln m)$ operations.

To prepare for the last step of the algorithm the n-tuple 1,2,3...n is then assigned to register C, leaving the computer in the state:

$$\frac{1}{n!}|\Psi\rangle \otimes \left(\sum_{\sigma \in S_n}|\sigma(1...n)\rangle\right)\otimes|1..n\rangle \qquad (3)$$

*Step IV. Sorting and unsorting.* The algorithm proceeds with a series of sorting and unsorting operations. A string of "scratch" qubits is required so that the sorting operations are reversible. Any sorting algorithm can be used; we suggest using a Heap sort, because it requires $O(n \ln n)$ operations in all cases and only $n \log_2 n$ scratch qubits. The first sort orders register B with a series of exchanges and scrambles A and C with the same series of exchanges. The resulting state is

$$\frac{1}{n!}\sum_{\sigma \in S_n}|\sigma(\Psi)\rangle|1..n\rangle|\sigma(1..n)\rangle|scratch\rangle \qquad (4)$$

At this point, one has already obtained a symmetrized superposition of the input states, but it is entangled with many other qubits. One can antisymmetrize simply by counting the number of exchanges made during the sorting operation and advancing the phase of that component of the superposition by $\pi$ if this number is odd [38]. The algorithm continues by reversing the sort on register B, but leaving registers A and C unchanged. The qubits contained in B and C are then redundant: in each component of the superposition, if B[i] = n, then C[n] = i. This redundancy allows B to be set to zero reversibly. By then sorting A and C together, eliminating C, and unsorting, one obtains the desired antisymmetrized state. Note that in the final unsorting operation, the algorithm relies upon the fact that the ordering of the input state $|\Psi\rangle$ was stipulated to be the same as the ordering of the integers 1..n in register C (so that antisymmetrization would be reversible); if this were not the case, the algorithm would fail. The entire process is completed in $O(n^2(\ln m)^2)$ operations.

Because the input state is now fully antisymmetrized, time evolution is in principle straightforward. Using the same technique as before, the Hamiltonian is split into a sum of terms $H_i$ and the corresponding time evolution operators $U_i = \exp[(-i/h) H_i (t/n)]$ are applied to the state in series. (The antisymmetry of the state will not be effected by truncation errors that occur during this process; although each individual $U_i$ does not preserve antisymmetry, their products do exactly. For a more detailed discussion of errors that occur during time-evolution, see [18].) Each $U_i$ can be performed by an appropriate series of quantum logic operations; the actual sequence of gates required can be determined by inverting the Campbell-Baker-Hausdorff formula. Using this procedure, $O(m^2)$ steps are required to perform an arbitrary one particle operator $U_i$, and $O(m^4)$ operations are required to perform an arbitrary two particle operator. It is therefore possible to simulate in polynomial time any system of fermions using the first quantized description (as long as the Hamiltonian does not include terms which involve k-particle interactions, where k is a large number of order n).

For the special case of the Hubbard model, the simplicity of the Hamiltonian allows one to perform each $U_i$ in only $O((\ln m)^2)$ steps. To begin, consider the Hubbard model Hamiltonian in its first quantized form:

$$H = \sum_{i=1}^{n} T_i + \frac{1}{2}\sum_{k \neq l} V_{kl} \tag{5}$$

where $\langle i\sigma|T|j\sigma\rangle = t_0 \delta_{<i,j>}$ and $\langle i\uparrow, i\downarrow|V|i\uparrow, i\downarrow\rangle = V_0$ are the only nonzero matrix elements of V. As before, the potential energy terms are easier because they are diagonal. For a given pair of particles, simply determine if they are at the same site and perform a controlled rotation if they are. In order to perform the time evolution corresponding to the kinetic energy terms, we focus on one particle at a time. For each particle, the idea is to decompose the kinetic energy terms into a sum of block diagonal matrices and then diagonalize the sub-blocks in each matrix in parallel [39]. For simplicity of explanation, we describe here only a 1-d Hubbard model and ignore spin. In this case the kinetic energy part of the Hamiltonian can be written

$$T = h(1,2) + h(2,3) + h(3,4) + ... + h(m-1,m) \tag{6}$$

where h(i,j) is the piece of the Hamiltonian that corresponds to hopping between sites i and j: all matrix elements of h(i,j) are zero except $\langle i|h(i,j)|j\rangle = \langle j|h(i,j)|i\rangle = t_0$. We rewrite the previous expression as follows:

$$T = T_1 + T_2$$
$$T_1 = h(1,2) + h(3,4) + h(5,6) + ... \quad (7)$$
$$T_2 = h(2,3) + h(4,5) + h(6,7) + ....$$

The operators $T_1$ and $T_2$ are in block diagonal form. To fully diagonalize each matrix (separately), perform quantum logic operations on each state to transform the state number into two quantum numbers labeling the block and the location within the block (0 or 1). For example, to diagonalize $T_1$, map the state $|n\rangle$ into $|(n+1) \text{ div } 2, n \text{ mod } 2\rangle$ (x div 2 indicates the greatest integer less than or equal to x/2). Because $T_1$ is block diagonal and only mixes states within each block - and because all states within the same block have their first quantum number in common - the action of $T_1$ takes place entirely within the space of the second quantum number. In this one qubit space it is simply the matrix $t_0 \begin{pmatrix} 0 & 1 \\ 1 & 0 \end{pmatrix}$. Thus all the blocks can be diagonalized in parallel by diagonalizing this trivial 2x2 matrix in the one qubit space of the second quantum number. Each state in the superposition is then advanced by the appropriate phase, and all the previous steps are reversed. This algorithm requires only $O((\ln m)^2)$ quantum logic operations.

Finally, we consider what information can be extracted from a quantum many-body simulation. It is obviously impossible to obtain the entire wavefunction: rather, the "answer" is obtained by performing a series of measurements on the qubits, one at a time. Each such measurement will yield either a $|0\rangle$ or $|1\rangle$. It is thus possible to measure any physical quantity that can be expressed in terms of such local variables. To obtain useful information about the physics of the simulated system, one must initialize the quantum computer, simulate time-evolution, make a measurement, and then repeat this process a sufficient number of times to acquire a statistically significant result. For example, the electronic charge density distribution can be obtained in the second quantized representation by performing measurements at each site to determine the probability of occupancy. The number of such measurements required to obtain some desired accuracy $\varepsilon$ varies as $\varepsilon^{-2}$ (i.e., the accuracy grows as a polynomial function of the number of trials). In the first quantized representation, the same result is obtained by measuring the location of a given particle and generating a histogram of locations from repeated trials. It is straightforward to obtain two-particle correlation functions and even k-particle correlations using a similar approach (requiring roughly $O(\varepsilon^{-2}\delta^k)$) trials, where $\delta$ is the density of points in the histogram and $\varepsilon$ is the desired accuracy). The momentum distribution function can be obtained by performing a quantum FFT before sampling the wavefunction. From the one and two particle densities and the momentum distribution, it is possible to calculate the system's expected energy. A variety of techniques can be used to obtain other information: for example, one can obtain scattering amplitudes by simulating the motion of an electron through a charged medium and measuring the probability of its emerging with different momenta. Or one could perform a quantum simulated annealing by time-evolving the system in contact with a simulated heat bath and then using the previous techniques to obtain information about the system's ground state. Thus one finds that, despite the unavoidable loss of information during the quantum measurement process, it is still possible to obtain desired information from the simulation, and that the process of doing so does not reduce the exponential improvements derived from using the quantum computer.

In summary, we have explicitly demonstrated how a universal quantum computer can be used to efficiently simulate systems consisting of many fermions. Depending on the particular problem, it may be preferable to employ second quantized notation (requiring m qubits) or first quantized notation (requiring $n \log_2 m$ qubits). A general algorithm for creating an antisymmetrized superposition of states has been described. We have also demonstrated algorithms which will simulate the Hubbard model, requiring $O(n^2(\ln m)^2)$ quantum logic operations in first quantized form, and $O(m^2)$ operations in second. The former algorithm employs a scheme for accommodating block diagonal Hamiltonians and could be applied to a wider range of problems.

D.S.A. acknowledges support from an NDSEG graduate fellowship and would like to thank J. D. Joannopoulos, Tomas Arias, and Steve Simon for helpful discussions. Portions of this research were supported through the DARPA initiative on Quantum Information and Computation (QUIC) administered by ARO.